\documentclass[sigconf,nonacm]{acmart}
\AtBeginDocument{%
  }

\usepackage{listings}
\usepackage{courier}
\usepackage{longtable}
\usepackage{verbatim}
\usepackage{graphicx}
\usepackage{subcaption}
\usepackage{bookmark}
\usepackage{tabularx}
\usepackage{tikz}
\usetikzlibrary{arrows.meta,positioning,calc}

\begin{document}

\title{The hidden life of signals: Time-domain inferences and other privacy attacks on everyday devices}
\titlenote{Presented in the Works-in-Progress track of the 2026 Hot Topics in the
  Science of Security Symposium (HotSoS 2026), April 14--16, 2026.
  Session details: \url{https://sos-vo.org/group/hotsos/2026/bratus};
  full agenda: \url{https://sos-vo.org/group/hotsos/agenda}. This is the
  authors' own version of the work, posted for non-commercial scholarly
  dissemination.}

\author{Larry Hernandez}
\orcid{1234-5678-9012}
\affiliation{%
  \institution{Dartmouth College}
  \city{Hanover}
  \state{New Hampshire}
  \country{USA}}

\renewcommand{\shortauthors}{Hernandez et al.}

\begin{abstract}
In privacy research on radiofrequency-based protocols, the dominant focus has remained on Bluetooth, WiFi, and
  Zigbee, while a broader and arguably more consequential attack surface has gone largely unnoticed: the privacy
  risks created by the \emph{composition} of everyday wireless protocols. Widely deployed systems such as KeeLoq
  remotes, vehicle TPMS sensors, and other sub-GHz devices continuously emit metadata and timing structure that, when
  analyzed jointly rather than in isolation, enable powerful behavioral inference. This work-in-progress paper argues
  that privacy leakage in these environments is not merely a property of individual protocols, but an emergent
  property of their interaction, correlation, and composition across devices, spaces, and routines. The resulting
  attack surface arises both from protocol metadata that directly degrades privacy and from the latent relationships
  between devices and the ways users move among and interact with them over time. We present preliminary evidence
  that these composed signals expose underappreciated opportunities for inference and tracking, and we outline a
  research agenda for characterizing and mitigating this broader class of privacy failures.
\end{abstract}


\maketitle

\section{Introduction}

Wireless communication protocols are ubiquitous in modern computing devices, ranging from
WiFi and Bluetooth to Zigbee, Z-Wave, and many others. These protocols facilitate interactions
between heterogeneous devices and their users. The security and privacy implications of these
protocols have been the subject of study, primarily focusing on the immediately obvious elements---such as device identifiers like MACs~\cite{acarPeekabooSeeYour2020}---and industry has responded by slowly remediating
and hardening the protocols against these issues.

For example, MAC address randomization (for IEEE 802.11/Wi-Fi) has been widely adopted \cite{fenskeThreeYearsLater2021,DBLP:journals/corr/MartinMDFBRRB17} to
mitigate tracking based on static hardware identifiers. This has rendered much---but not all~\cite{vanhoefWhyMACAddress2016}---of the published
research obsolete, except for specialty attacks relying on timing and RF characteristics \cite{5544294},
under laboratory controlled conditions.%

Similarly, Bluetooth Low Energy (BLE) privacy features have been
implemented to mitigate the same kinds of tracking. While other methods of de-anonymization exist \cite{dasUncoveringPrivacyLeakage2016, beckerTrackingAnonymizedBluetooth2019, martinHandoffAllYour2019},
these are typically specific to the modulation and physical layer characteristics of the
protocols and devices such modems and baseband chips~\cite{joverLTESecurityProtocol2016}, and are neither generalizable nor
scalable, requiring highly specialized hardware and software. These emergent properties of modulation
have also been successfully leveraged to create unexpected
capabilities such as a Wi-Fi-based radar within confined spaces \cite{li2020taxonomy,li2020passive}.

Nevertheless, more subtle interactions between the less explored wireless protocols~\cite{houCloakLoRaCovertChannel2020} and their abuses to degrade user privacy remain largely unexplored. This includes protocols
such as Keeloq~\cite{microchip_hcs301_ds21143c} (used in remote keyless entry systems such as vehicles and garage doors), the ubiquitous
Tire Pressure Monitoring System (TPMS)~\cite{grygier2001evaluation}, and many others. These protocols often contain metadata
and design choices that can be readily exploited to infer user behavior, device usage patterns
(pattern-of-life), and other sensitive information.

It is these interactions and manifestations of user behavior intersecting with protocol-specific
characteristics, that present a tremendous---and mostly unexplored---attack surface.

\subsection*{\bf Our contributions}

In short:

\begin{itemize}
  \item A multi-year empirical dataset of sub-GHz RF activity across heterogeneous protocols.
  \item Demonstration that time-domain correlation alone (no PHY fingerprinting, no decryption) enables behavioral inference.
  \item Evidence that cross-protocol composition materially increases privacy leakage relative to single-protocol analysis.
  \item Identification of protocol design features (cleartext metadata, repetition semantics) that systematically enable inference.
\end{itemize}

We present preliminary results from ongoing research into the privacy implications of {\bf composition and
  interactions} of%
various wireless protocols, focusing on time domain inferences and protocol-specific attacks. We outline future
directions for this line of research, including analysis of the interactions between different
protocols, including how clustering and correlation techniques can be used to infer sensitive information,
and the challenges in mitigating these privacy risks.

Previous work has focused on individual protocols, such as Wi-Fi, BLE \& Bluetooth, in isolation, and often limited to stationary devices---such as ZigBee---while often neglecting the broader
ecosystem of wireless communications and the interactions between different protocols and devices. More
importantly, prior work was typically oblivious to how users actually interacted with the devices employing these protocols. Our work 
fills this gap by exploring the complex relationships between various wireless protocols and
user behavior, shedding light on \emph{the hidden life of signals} and their implications for user privacy.

\section{Under-researched yet widely deployed RF protocols}
\label{s:rf-protocols}

While much of the research in wireless protocol security and privacy has focused on popular protocols
in the 2.4 GHz ISM band, many other protocols operate in the sub-GHz ISM bands (e.g., 315 MHz, 433 MHz).
The modulation schemes in WiFi and Bluetooth enable specific attacks and side-channels, but these are
neither applicable nor present in other modulation schemes, such as OOK, ASK, FSK, etc. Prior work related to
Wi-FI has predominantly focused on:

\begin{itemize}
  \item Passive radar capabilities \cite{li2020taxonomy,li2020passive,chetty2011through} relying on protocol-specific
  artifacts such as CSI or modulation artifacts derived from the use of OFDM \cite{li2020passive};
  \item Privacy leaks via protocol-level identifiers such as MACs, SSIDs, and other management frame information;
  \item Timing and payload variance analysis \cite{acarPeekabooSeeYour2020}.
\end{itemize}

Notably, protocol-level {\em interactions} between Wi-Fi and wired Ethernet seem overlooked.
For instance, as Wi-Fi access points effectively behave as network bridges, they must translate broadcast
traffic between the two media. MAC addresses from the wired side of the network are often leaked through
Access Points (APs) as part of their management and broadcast frames, allowing a passive observer to infer and potentially
reconstruct wired network topologies and device presence, such as presence of specific upstream switching
or routing infrastructure. The latter infrastructure elements are highly visible as they seldom, if ever, change their MAC addresses.

Beyond Wi-Fi and Bluetooth, other 2.4GHz ISM band protocols include Zigbee, Thread, and Z-Wave. These protocols
have been the subject of research focusing on their security vulnerabilities and resilience against attacks
specialized against mesh networks. Again, the privacy implications at the intersection of human behavior---interaction with the devices---and protocol characteristics remain largely unexplored.

Finally, other sub-GHz ISM band protocols ---such as Keeloq and TPMS--- have received little attention.
These protocols are massively deployed, in the number of billions of devices, hard or impossible to
update or patch, and often contain meta-data that can be exploited to infer sensitive information about user
behavior. While PHY level characteristics of these protocols may not offer the opportunities for passive
radar and similar techniques, time-domain characteristics of their utilization patterns, in conjunction
with protocol-specific meta-data, present a ubiquitous and poorly understood attack surface.

LoraWAN \cite{haxhibeqiri2018survey,yang2018security} and SigFox \cite{lavricSigFoxCommunicationProtocol2019} are
two additional protocols seeing increased adoption in IoT deployments. SigFox is marketed as a resilient
protocol for alarm systems in high-risk contested environments ---where jamming of the local centralized unit and its
sensors should be ideally detected and reported to a remote monitoring station. These protocols are already entering
the residential and commercial security market, while their privacy implications remain unexplored.

Arguably even more concerning is the use of proprietary and undocumented protocols in critical infrastructure.
This includes TETRA and TETRAPOL \cite{10.1108/ICS-12-2024-0318,esmaeilifar2025public}, which also operate in the sub-GHz ISM bands, and are widely used in public safety,
military, and emergency services communications~\cite{esmaeilifarPublicSafetyNetworks2025,wursterDevelopmentPublicSafety2013}. Security and privacy implications of these protocols
has seen limited research, especially with respect to privacy implications of metadata analysis in
the context of passive interception \cite{meijer2023all,pfeifferAnalyzingTETRALocation2016}.

\section{Collection and analysis}

\subsection{Experimental setup}
\label{sec:experimental-setup}

Our experimental setup was designed to capture and analyze RF signals from ISM-band devices
across various protocols, including Keeloq, TPMS, and several others operating in the
315--434 MHz frequency range.

Our setup consists of the following components:

\begin{itemize}
  \item A software-defined radio (SDR) platform (HackRF One and LimeSDR) capable of capturing a wide range of frequencies and modulation schemes.
  \item A set of antennas optimized for the 315 MHz and 433 MHz ISM bands to ensure efficient signal reception.
  \item A low noise amplifier (LNA), set of bandpass filters, and other RF front-end components to distribute and enhance signal quality and reduce interference.
  \item An embedded ARM64 computer running GNU Radio, \path{rtl_433} \cite{larssonMerbananRtl_4332026} and custom signal processing software for real-time channelizing, signal analysis and demodulation.
  \item A time-series database (InfluxDB) backend for storing captured signals and their metadata for further analysis.
  \item Protocol decoders from \path{rtl_433} \cite{larssonMerbananRtl_4332026}.
\end{itemize}

Capabilities for 2.4 GHz ISM band collection were also separately configured but are not covered
in this paper due to time and space constraints. The principles however remain equally applicable
to protocols in that frequency range.

The setup was designed for minimal cost, maximum portability, and maximum ease of deployment in
various environments. The antennas chosen are compact dipole designs with adhesive backing that can
be readily installed in false ceilings and walls, allowing for discreet signal capture in
residential and commercial settings. Low heat dissipation components were selected to enable
long-term operation without active cooling, making the setup suitable for extended monitoring. This
enabled collection in a multi-year span.

Two locations were employed for data collection, chosen for their typical device density and diversity,
providing a rich dataset for analysis, and the permissive jurisdiction for non-invasive passive
signal collection.

\subsubsection{Site A}
\label{siteA}

A residential setting exposed to transient and stationary vehicle and mobile device traffic
with the aforementioned configuration of low-profile antennas and receivers.

\subsubsection{Site B}
\label{siteB}

A licensed antenna installation with city-scale visibility, employing three colinear antennas
for VHF/UHF communications.

\subsection{Ethics and scope}

Data collection was conducted without interaction or intervention involving individuals,
with no intentional identification of persons and no linkage to personal identity.
All signals analyzed were radiofrequency emissions broadcast by design, and no encryption,
authentication mechanisms, or access controls were bypassed. Collection therefore
mirrored established spectrum-monitoring and passive measurement practices\footnote{All observations were limited to license-free ISM bands, whose reception and monitoring
require no licensing or reporting obligations in the jurisdictions where data was collected.}.

Data was analyzed only after the conclusion of the collection period and otherwise left
unattended. No attempts were made to deanonymize devices, users, or locations.
Analysis was performed in aggregate and pseudonymous form, reflecting the absence
of any linkage to personal, household, or property identity.

\subsection{Data metrics}
\label{sec:data-metrics}

As of January 2026, the dataset contains over $1.6$ million unique records, beginning in August 2023,
of which:

\begin{itemize}
  \item Over 116337 records correspond to successfully demodulated and decoded TPMS sensor signals. These contain
  11,984 unique sensor IDs and 547 unique vehicle associations inferred from sequentiality of the identifiers. Prior to
  grouping by vehicle, a total of 243 sensor IDs were observed more than 50 times.
  \item 352 unique IDs were observed in the Keeloq remote keyless entry signals, successfully decoded
   as HCS301/HCS200 \cite{microchip_hcs301_ds21143c} protocol.
  \item The distribution of captured signals is highly skewed and long-tailed, displaying clear dominance of
  TPMS and EV1527 / PT2262-like rolling/learning code remotes, as shown in Table~\ref{tab:dataset_top_typed_confidence}.
\end{itemize}

Table~\ref{tab:dataset_top_typed_confidence} summarizes the top measurements by record count,
along with inferred device types and false-positive likelihood assessments based on protocol characteristics
and known device behaviors. Table~\ref{tab:tpms_measurements_rows} lists the top TPMS measurements by
record count.

\subsubsection{Grouping and attribution}

Decoding simple OOK/FSK modulated protocols successfully does not guarantee accurate data recovery unless
error correction mechanisms are in place\footnote{The presence of CRC fields is far more extended than proper error recovery codes.}. This is especially relevant to fields like identifiers, mostly
numeric, which are crucial for grouping and attribution of signals to specific entities (e.g., vehicles, remotes, and
sensors).

In the case of TPMS, the decoding process includes CRC-8 checksums that validate the integrity of the
payload, yielding a high confidence in the integrity of the decoded bitstream. This facilitates 
grouping and observations, for example with observed sequential IDs in TPMS sensors, suggesting that
manufacturers assign IDs in a sequential manner during production ---a likely side-effect from
assembly line processes. Whether this behavior is consistent across manufacturers and models is
unclear, but it provides another dimension for actionable attribution: if the statistical variance
for a specific manufacturer ---or year of production, or model--- is known, this could be used for
targeted tracking or identification of vehicles based on the sequentiality or variance patterns
of their TPMS IDs. This would be fragile to after-market TPMS replacements, but could still provide
valuable information in certain scenarios.

\begin{figure}[h]
  \centering
  \includegraphics[width=\linewidth]{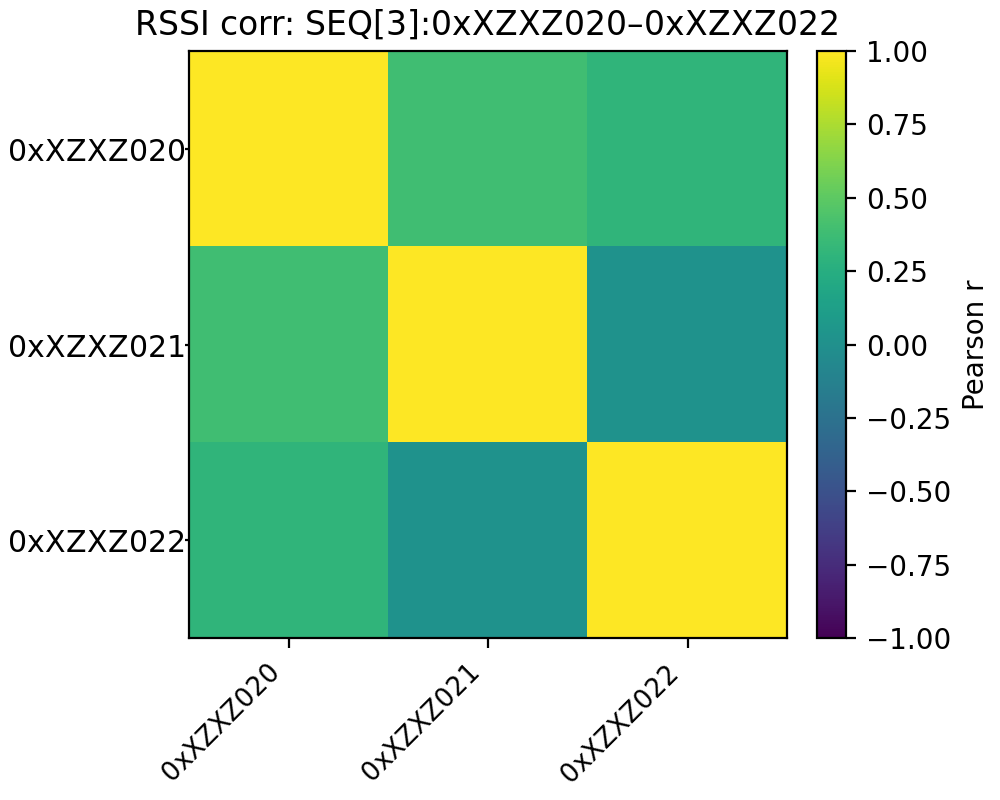}
  \caption{Example RSSI correlation of a specific vehicle and a set of four TPMS sensors with sequential IDs}
  \Description{XXX}
  \label{tab:tpms-group-headmap}
\end{figure}

In practice, a hybrid approach was employed to correlate TPMS sensors through their numerical
in-protocol identifiers, using sequentiality and RSSI values---arguably not reliable on its own for
these signals, but effective in composition with other data. A minimal set of two IDs was sufficient for meaningful grouping.
Temporal correlation based off IDs consistently provided the most reliable grouping. RF receiver
sensitivity and better antenna systems will drastically improve collection. Because of the periodicity,
frequency and movement patterns, Doppler effect does not seem like a viable or particularly
interesting metric to factor in.

\begin{table*}[t]
\footnotesize
\centering
\begin{tabularx}{\textwidth}{@{}l r l X X@{}}
\toprule
\textbf{Measurement (Human)} & \textbf{Rows} & \textbf{Type} & \textbf{Note} & \textbf{FPs likelihood} \\
\midrule
Mixed Weather Sensor and OOK/FSK Misclassification  & 713{,}650 & Alarm or alarm sensor &
Consumer weather station telemetry. &
Likely generic OOK/FSK weather-sensor protocol (e.g. GSDWS07); may include multiple vendors using similar RF ICs (e.g., low-cost TX ASICs or PT2262-class encoders). \\

Kerui Alarm Sensors & 381{,}705 & Alarm or alarm sensor &
Home alarm entry and PIR sensors. &
\\
Microchip HCS200-300 (KeeLoQ) & 229{,}264 & Rolling code remote (HCS200, HCS301) &
KeeLoQ rolling-code encoder IC. &
 \\

Toyota TPMS (Vehicle) & 42{,}928 & TPMS &
Automotive wheel pressure sensors. &
 \\

Jansite TPMS (Vehicle) & 42{,}230 & TPMS &
Aftermarket TPMS ecosystem. &
 \\

PT2262/PT2272 or EV1527-class & 18{,}079 & Rolling/learning code remote &
Consumer gate or garage remote. &
Likely PT2262/PT2272-like encoder or EV1527-class IC; similar remotes (e.g.  Akhan 100F14 Remote) may be misattributed \cite{EV1527SubdecodersAkhan100F14}. \\

Honda Car Remote & 15{,}426 & Car remote &
Automotive key fob frames. &
 \\

Smoke GS558 Detector & 14{,}711 & Alarm or alarm sensor &
RF smoke detector sensors. &
 \\

Renault TPMS (Vehicle) & 13{,}229 & TPMS &
Automotive wheel pressure sensors. &
\\

Regency Remote & 5{,}533 & Learning code remote &
Consumer gate or garage remote. &
Likely EV1527 or compatible encoder family. \\

Hyundai VDO TPMS (Vehicle) & 4{,}404 & TPMS &
Automotive wheel pressure sensors. &
 \\

Abarth 124 Spider TPMS (Vehicle) & 3{,}189 & TPMS &
Automotive wheel pressure sensors. &
\\

Markisol Shade Remote & 2{,}978 & Learning code remote &
Motorized shade or blind control. &
Often based on EV1527/PT2262-class encoders; many blind remotes share identical RF protocols. \\

Ford TPMS (Vehicle) & 2{,}484 & TPMS &
Automotive wheel pressure sensors. &
\\

Cardin S466 Remote & 2{,}482 & Rolling/Learning code remote &
Gate or garage automation remote. &
Cardin remotes often reuse common rolling-code encoder families. \\

Emos TTX201 Remote & 2{,}122 & Rolling/Learning code remote &
Consumer control transmitter. &
 Likely EV1527/PT2262-class IC (attribution may overlap). \\

Schrader TPMS Sensor Family (EG53MA4) & 2{,}013 & TPMS &
Schrader automotive TPMS sensors. &
 \\

Schrader TPMS Sensor Family & 1{,}931 & TPMS &
Schrader automotive TPMS sensors. &
\\
Renault 0435 R TPMS & 1{,}561 & TPMS &
Renault-associated TPMS variant. &
\\

HT680 Remote & 1{,}549 & Rolling/Learning code remote &
Consumer gate or garage remote. &
Likely generic rolling/learning-code encoder family. \\

Citroen TPMS (Vehicle) & 1{,}378 & TPMS &
Automotive wheel pressure sensors. &
 \\
\bottomrule
\end{tabularx}
\caption{Top signal types by record count with inferred device type and relevant context}
\label{tab:dataset_top_typed_confidence}
\end{table*}

\subsection{Clustering analysis}

The data was retrieved from InfluxDB and converted into Parquet format for efficient storage and retrieval. Initial
exploratory data analysis was performed using Python's Pandas and NumPy libraries, allowing for
statistical analysis and visualization of the captured signals. Clustering algorithms, such as
K-Means and DBSCAN, were applied to identify patterns and group similar signals based on their
time domain characteristics. Feature extraction techniques, including time domain statistics,
were used to derive meaningful attributes from the raw signal data, facilitating the clustering
process. Visualization tools, such as Matplotlib, were employed to create plots and
charts.

Record volume by decoded measurement (infrastructure rows excluded).
Horizontal axis uses logarithmic scaling with linear count labels to preserve
visibility across the long-tailed distribution.

\section{Preliminary results}

\subsection{Challenges in decoding}
\label{sec:decoding-challenges}

Certain ICs of rolling code remotes ---such as the EV1527 and PT2262 families--- exhibit significant
variance in how the device encodes the transmitted data. This variance reflects as
false positives in the decoding process, as the same protocol may be implemented differently across
different manufacturers and models. For example, the EV1527 family can often be classified as Akhan 100F14 remotes.
This is a known limitation of the rtl433 software~\cite{EV1527SubdecodersAkhan100F14}, which relies on community-contributed decoders
and heuristics to identify and decode signals. As such, the attribution of decoded signals to specific
devices in these cases becomes considerably less reliable. An approach to mitigate this issue involves
perusing a library of known signals confirmed to belong to specific devices, implicitly requiring reverse
engineering of their modulation and encoding schemes.

\subsection{Time-domain inferences of vehicle and household signals}

The inherent nature of how some of these devices are used carries privacy implications and
observable metrics and traits clearly visible in the time domain with high correlation to
user behavior. For example, vehicle TPMS sensors~\cite{roufSecurityPrivacyVulnerabilities,arduino_forum_renault_tpms_2025} transmit data only when the vehicle is in motion,
and the frequency and timing of these transmissions can be used to infer vehicle usage patterns,
including trip start and end times, duration, and frequency. Similarly, remote keyless entry systems
transmit signals when the user interacts with the vehicle, such as locking or unlocking doors. These interactions
can be correlated with other signals, such as entry sensors in residential settings, to build a comprehensive picture of user behaviors and routines.

As such, the time domain characteristics of these signals can be exploited to infer sensitive information about users,
even in the absence of payload decryption or protocol-specific knowledge. This highlights the need for
a deeper understanding of the privacy implications of these protocols and the development of mitigation strategies
to protect user privacy.

\begin{figure}[h]
  \centering
  \includegraphics[width=\linewidth]{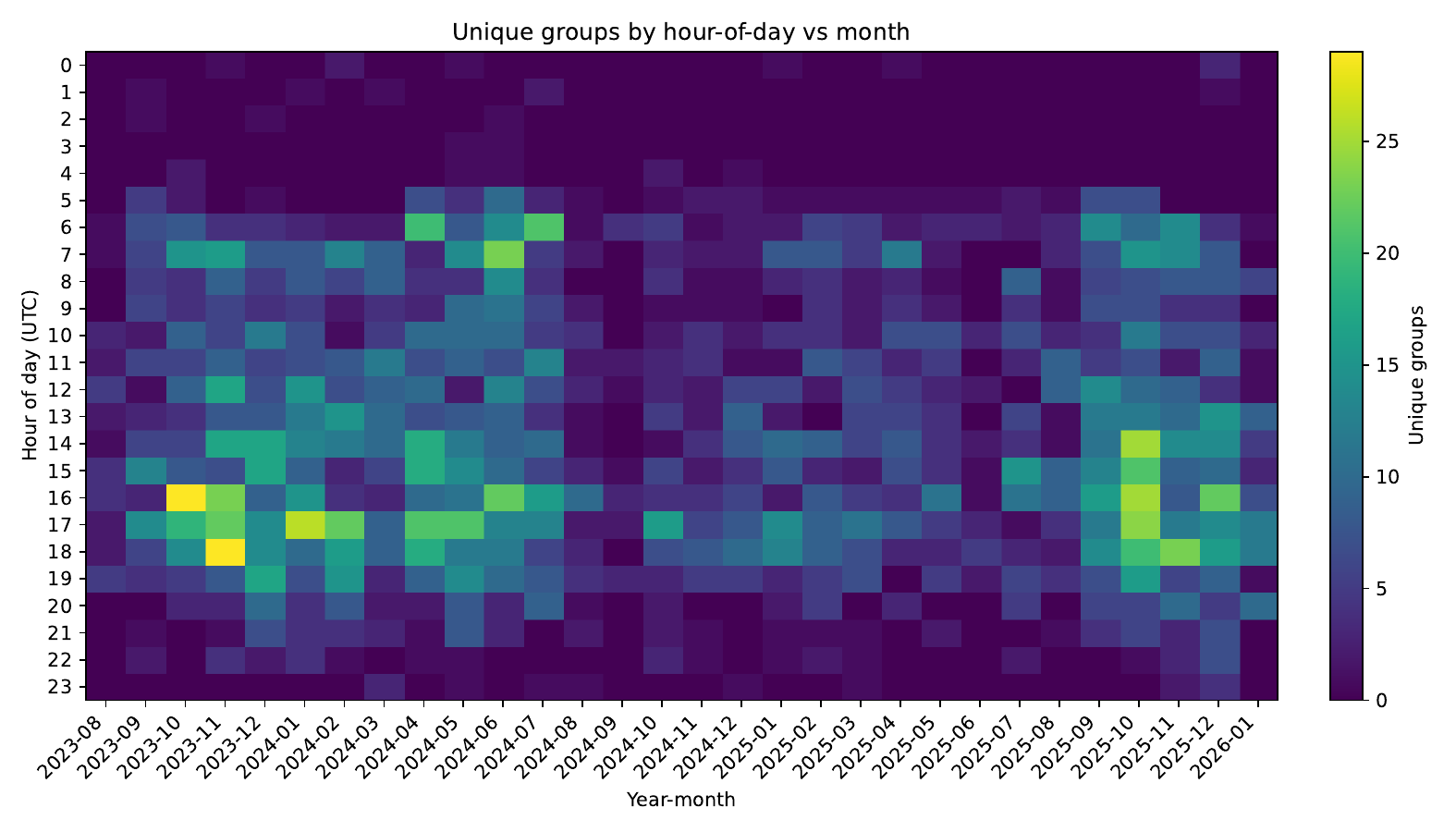}
  \caption{Heatmap of collected signals from TPMS sensor groups (2023 Aug-2026 Jan)}
  \Description{The figure displays a heatmap of the correlated TPMS sensors inferred through timing and ID analysis,
  in the span of multiple years between 2023 and 2026.}
  \label{fig:tpms-group-headmap}
\end{figure}

Table~\ref{fig:tpms-group-headmap} displays the heatmap of presence of clustered TPMS sensor groups
across multiple years, with month granularity. Clear observable patterns include reduced presence
during summer months, and much higher density during early morning and late afternoon hours, corresponding
to regular office hours.

\subsection{Cross-protocol correlations and inferences}

Behavioral interactions between the different devices and their users  create a rich ``hidden life''
patterns of their signals, where the timing and sequences of events across multiple protocols can be correlated
to infer sensitive information about user behaviors and routines.

Oftentimes, these interactions also reflect the physical layout and configuration of the user's environment.
For example, the metadata in Keeloq signals from remotes carries both identifier and button press information.
In residential and business settings, for convenience and cost, one remote with multiple buttons can have each
button associated with a distinctive receiver, dependent on support by the receiver and employed IC. This is
the case for HCS301-based systems. As such, the button metadata will reveal both the number of distinct
perimeter access points (e.g., garage door, outside gate, vehicle trunk, etc) and the timing of their use.
In correlation with TPMS signals from vehicles, home alarm sensors and systems, which can provide information about entry, exit, and movement patterns within the property. Thus, a detailed picture of user behaviors and routines can be constructed.

The ability to cluster and analyze these signals based on their unique identifiers can also provide insights
into the behavior of multiple users within the same vicinity. For example, when the same remote is utilized to
manipulate multiple perimeter access points (e.g., garage door and outside gate), the correlation between these
events can be established, revealing patterns of movement and access within the property, such
as arrival and departure times, dwell time within specific perimeters, vulnerable patterns (e.g. leaving an outside
gate open for extended periods, or at the same time another perimeter gate is left open) and other
``pattern of life'' patterns of sensitive information.

\subsection{Case study: Keeloq and vehicle signals}

The Keeloq protocol \cite{microchip_hcs301_ds21143c}, widely used in remote keyless entry systems for vehicles and garage doors,
contains distinct identifiers in its non-encrypted payload that can be exploited for clustering
and analysis. More specifically, it contains a 28-bit serial number in the cleartext portion of its
transmitted word (see Table~\ref{tab:hcs301_singlecol}). This serial number uniquely identifies
the transmitter device (tables ~\ref{tab:hcs301_singlecol} and \ref{tab:hcs301_enc_fields}), allowing for the grouping of signals originating from the same source.
Additionally, it contains a button identifier that indicates which button was pressed
(e.g., lock, unlock, trunk release). This button identifier is also transmitted in the clear-text
portion of the message, providing further granularity for analysis. By analyzing the timing and frequency of these button presses,
it is possible to infer user behavior, such as when the vehicle is being accessed, how often it
is used, and potentially even the user's routine. For Keeloq-based garage door openers, similar inferences can be made
regarding when the user arrives home or leaves, based on the timing of the door (receiver) opening and closing events.

\begin{figure}[h]
  \centering
  \includegraphics[width=\linewidth]{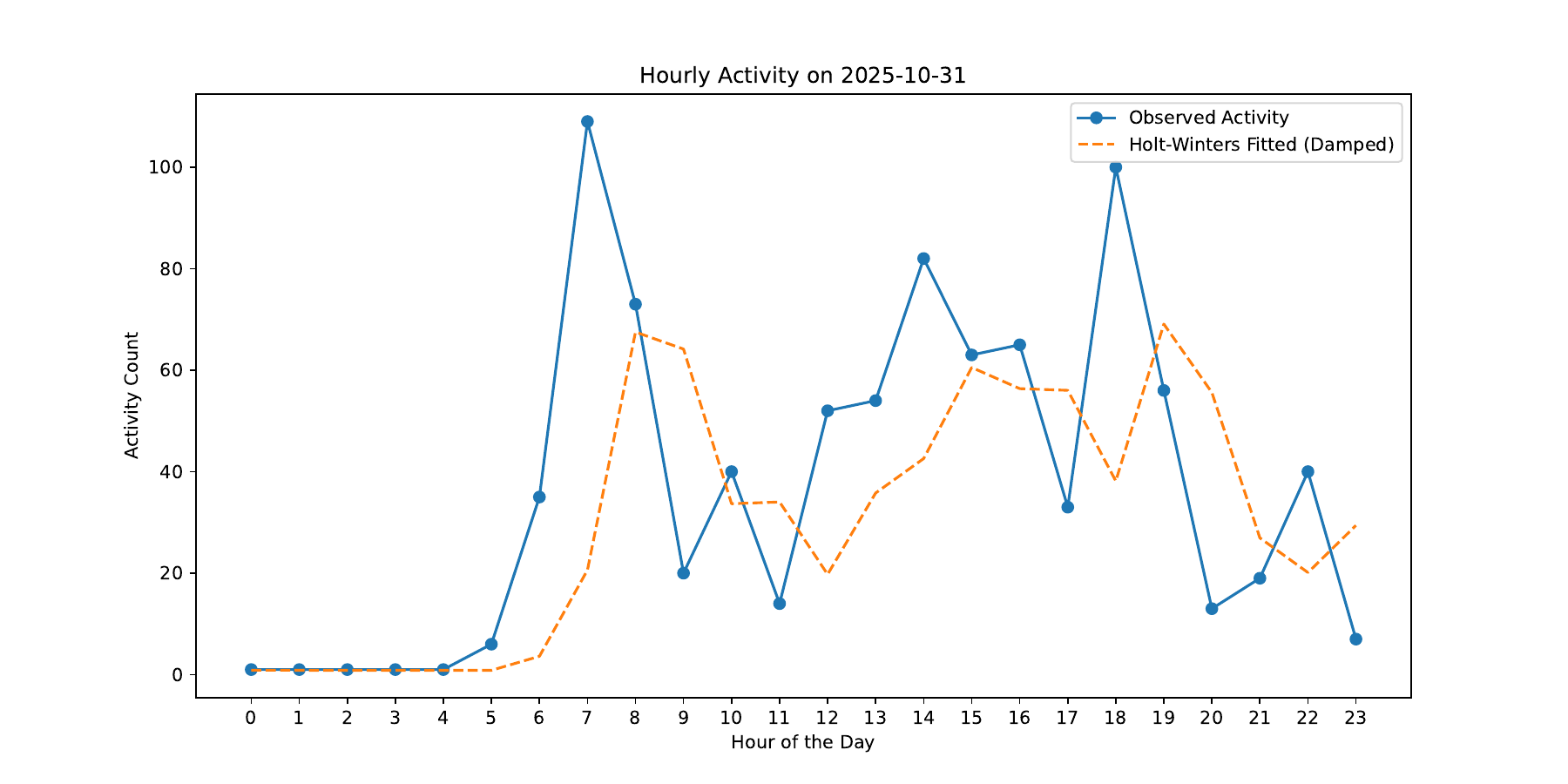}
  \caption{Uniquely identified Keeloq activity within a 24-hour span}
  \Description{XXX}
  \label{fig:keeloq-daily-activvity}
\end{figure}

Figure~\ref{fig:keeloq-daily-activvity} shows the collected Keeloq signals distribution across 24 hours
---daily activity---, displaying overlapping clustering with TPMS sensors as shown in Figure~\ref{fig:tpms-group-headmap}.

\subsection{Keeloq button clustering times}

The clustering of Keeloq button presses based on their time domain characteristics revealed
distinct patterns corresponding to different user interactions with distinct gates and vehicles. For example, the clustering analysis identified
groups of button presses that occurred in close temporal proximity, suggesting that they were associated with the same user action, such as unlocking a vehicle and opening a garage door
upon arrival at home. These clusters were characterized by short inter-press intervals of different buttons of the same remote,
typically within consistent time windows (e.g., within a few seconds of each other).

\begin{figure}[h]
  \centering
  \includegraphics[width=\linewidth]{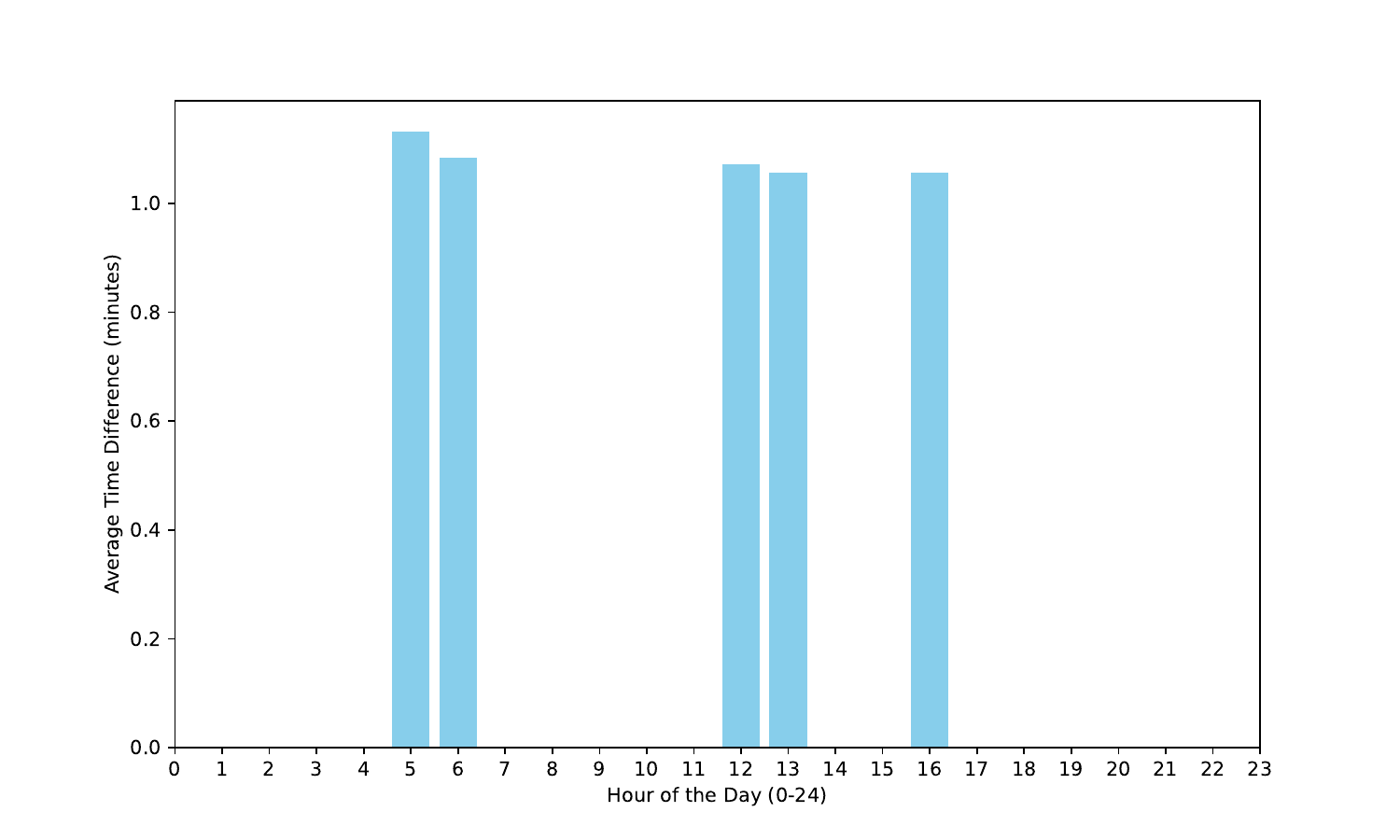}
  \caption{Distribution of distinct unique Keeloq remote intra-button presses, with dwell time between buttons}
  \Description{XXX}
  \label{fig:keeloq-button-window}
\end{figure}

Figure~\ref{fig:keeloq-button-window} displays the temporal distribution ---histogram--- of
unique remotes with distinctive button presses with a maximum of 5 (five) minutes between each
button press. These signals from Site A (Section~\ref{siteA}) display the convergence in time-domain
of user behavior, with clear clustering of Keeloq and TPMS signals. Similarly, other protocols
involved in alarm systems (e.g. motion sensors, etc) contain similar manifestations of proximal
activity. In the case of simple 433 MHz alarm sensors, the motion data is transmitted openly,
and is likely that it extends to MWave radar sensors (typically Zigbee), for example. This
enables inference of movement and device interactions within the property, across multiple
heterogenous events and devices.  Inference can be extended to determine when the user operates
the trunk of the vehicle, when they leave a gate open ---before another perimeter closes---, if
they choose to arm their security system in ``stay'' mode ---or not---, room occupancy, and
so forth.

\section{Discussion and future work}

Planned future work includes expanding the dataset (Section~\ref{sec:data-metrics}) to include a wider variety of devices and protocols,
as well as refining the clustering algorithms to improve the accuracy and granularity of the analysis. Additionally, we plan to explore
the development of mitigation strategies to protect user privacy, such as protocol modifications or
the implementation of privacy-preserving techniques. Furthermore, we plan to investigate the potential for real-time monitoring and analysis of these signals,
enabling proactive detection of privacy-invasive behaviors and the development of practical
countermeasures that account for the radiofrequency characteristics of the different protocols and
their constraints, such as frequency hopping or duty cycling (channel occupation and ``flight time'').

Additionally, refinement of the experimental setup to enhance signal capture quality and reduce interference
is planned. This includes the use of site-specific directional antennas, improved RF front-end components, and
custom software optimizations for real-time signal processing, including FPGA-based demodulation
and decoding for the common protocols using platforms such as the Pluto and ANT-SDR hardware.

\subsection{What will not work}

Countermeasures based on fabricated or decoy transmissions are unlikely to be effective:

\begin{itemize}
  \item The timing, structure, and repetition characteristics of fabricated signals would
  need to closely mimic those of genuine devices. Any solution capable of such fidelity
  would incur substantial engineering complexity and cost, far outweighing its practical
  benefits in this context.
  \item Channel occupation constraints at the physical layer limit the feasible volume and
  frequency of fabricated transmissions. Injecting decoy signals into shared spectrum—
  particularly in the already congested and fragile 315/433\,MHz ISM bands—risks degrading
  legitimate communications and effectively creating a self-inflicted jamming scenario.
\end{itemize}

In general, such \emph{ad hoc} patchwork solutions are not only ineffective, but also divert
engineering effort away from more robust and systematic approaches, including protocol
redesign and the incorporation of privacy-preserving mechanisms at the protocol level.

Mitigating this class of vulnerabilities requires protecting all protocol metadata that is
meaningful to a passive adversary. Where feasible, protocols should additionally incorporate
defenses against payload timing analysis. Such measures are likely impractical for protocols
based on simple OOK modulation ---due to direct exposure of symbol timing and airtime
competition if employing decoy transmissions---, but may be more attainable for FSK-based designs
because of inherent tolerance of randomized padding, variable preambles, et cetera. Still,
these changes carry implicit engineering costs, before even reaching the market adoption
and deployment stages. In short, the re-design and manufacturing of completely new semiconductors
(ICs) and their adoption is an extraordinary and extraordinarily lengthy process.

\subsection{Mitigation at scale and device obsolescence}

The devices most affected are often those with long lifespans and limited update capabilities,
such as vehicles and certain IoT devices. Mitigating these privacy risks at scale inherently
involves very significant supply chain and manufacturing challenges, as well as user adoption
barriers. Many of these devices are not designed to be updated or modified post-deployment,
making it difficult to implement changes that would enhance privacy protections.

The rate of obsolescence or replacement of devices such as garage doors, TPMS sensors, and keyless
entry mechanisms, has not been adequately studied in literature  or industry. In essence,
many of these devices are ``install and forget,'' and this extraordinary lifespan is precisely what makes them most
vulnerable and attractive to adversarial exploitation.

\subsection{Ground truth}

The difficulty in establishing ``ground truth'' in this context is the obvious and arguably 
unsurmountable challenge in establishing a multi-residence, multi-individual, multi-vehicle,
multi-device ecosystem, where individuals log and document their every interaction with each
and every device producing a measurable or collectable RF signal. This would necessitate
of full, permanent cooperation, if only limited initially to a subset of devices.

The more realistic and approachable challenge of course is to ensure the reproducibility of the experimental setup, 
the availability of the data under fair use causes, respectful to all ethical considerations,
and the careful assessment of all potential pitfalls (Section~\ref{sec:decoding-challenges}).

\begin{acks}
I would like to thank my advisor and friend, Prof. Sergey Bratus for his support,
motivation and odd-hours review efforts with this paper. I also extend my gratitude
to my peers and friends in Europe as they maneuver difficult times, to whom I owe 
much of my knowledge in this domain.
\end{acks}

\bibliographystyle{ACM-Reference-Format}
\bibliography{hidden-life-of-signals}

\newpage

\appendix

\section{Additional figures and tables}

\subsection{TPMS dataset metrics}

\begin{table}[h]
\footnotesize
\centering
\begin{tabularx}{\columnwidth}{@{}l c@{}}
\toprule
\textbf{Measurement} & \textbf{Total} \\
\midrule
Toyota             & 42{,}928 \\
Jansite            & 42{,}230 \\
Renault            & 13{,}229 \\
Hyundai-VDO        & 4{,}404  \\
Abarth-124Spider   & 3{,}189  \\
Ford               & 2{,}484  \\
Schrader-EG53MA4   & 2{,}013  \\
Schrader           & 1{,}931  \\
Renault-0435R      & 1{,}561  \\
Citroen            & 1{,}378  \\
\bottomrule
\end{tabularx}
\caption{Top TPMS signals (FSK) by successfully decoded record count}
\label{tab:tpms_measurements_rows}
\end{table}

\subsection{Microchip HCS301 payload structure}
\begin{table}[h]
\footnotesize
\centering
\begin{tabularx}{\columnwidth}{@{}l c X@{}}
\toprule
\textbf{Field} & \textbf{Bits} & \textbf{Description} \\
\midrule
Encrypted block & 32 &
KeeLoQ-encrypted payload containing button code, discrimination bits,
and 16-bit synchronization counter. \\
Fixed block & 34 &
Cleartext fields used for receiver identification and filtering. \\
\midrule
Status & 2 &
Low-voltage and repeat/status indicators. \\
Button (clear) & 4 &
Button identifier transmitted in the clear. \\
Serial number & 28 &
Unique transmitter identifier. \\
\bottomrule
\end{tabularx}
\caption{HCS301 transmitted word structure (66 data bits, LSb first)}
\label{tab:hcs301_singlecol}
\end{table}

\begin{table}[h]
\footnotesize
\centering
\begin{tabularx}{\columnwidth}{@{}l c X@{}}
\toprule
\textbf{Sub-field} & \textbf{Bits} & \textbf{Role} \\
\midrule
Button & 4 &
Participates in cipher input; must match cleartext button field. \\
Discrimination & 12 &
Receiver-side prefilter to reduce key search space. \\
Sync counter & 16 &
Monotonic counter providing replay protection. \\
\bottomrule
\end{tabularx}
\caption{Logical composition of the 32-bit encrypted portion of an HCS301
code word.}
\label{tab:hcs301_enc_fields}
\end{table}

\subsection{TPMS payload structures}

\begin{table}[h]
\footnotesize
\centering
\begin{tabularx}{\columnwidth}{@{}c l X@{}}
\toprule
\textbf{Byte(s)} & \textbf{Field} & \textbf{Interpretation} \\
\midrule
0 & Flags + Pressure[9:8] &
Upper bits encode status flags; lower bits carry MSBs of pressure. \\
1 & Pressure[7:0] &
Remaining pressure bits; scale 0.75 kPa per LSB. \\
2 & Temperature &
Signed Celsius value with offset of $-30^\circ$C. \\
3--5 & Sensor ID &
24-bit little-endian unique sensor identifier. \\
6--7 & Reserved &
Observed as constant or model-dependent values. \\
8 & CRC-8 &
CRC-8 checksum (poly 0x07, init 0x00). \\
\bottomrule
\end{tabularx}
\caption{Renault TPMS payload format (9 bytes)}
\label{tab:renault_tpms_singlecol}
\end{table}

\subsection{RF block diagrams}

The design of the RF front end used in the experimental setup is illustrated in
Figure~\ref{fig:rf-block}. The design incorporates dual-band coupling to facilitate
simultaneous reception of signals in the 315 MHz and 433 MHz ISM bands. This is achieved through the use of
a bandpass filter for each frequency band, followed by a low noise amplifier (LNA) to
compensate for losses, prior to distributing the signals to two receivers. Alternative
designs are provided for a single receiver using a combiner again, or an upconversion
step (translating the 315 MHz signals into the 433 MHz range). It must be noted that
most 433 MHz signals are actually centered at 433.92MHz, but most SDR platforms will
be able to gracefully detect the peak and adjust to either.

\begin{figure*}[t]
  \centering

\begin{tikzpicture}[
  font=\small,
  >=Latex,
  block/.style={draw, rounded corners, minimum width=10mm, minimum height=10mm, align=center},
  rf/.style={block},
  sdr/.style={block, minimum width=30mm},
  ant/.style={draw, minimum width=18mm, minimum height=8mm, align=center},
  coupler/.style={draw, rounded corners, minimum width=20mm, minimum height=18mm, align=center},
  line/.style={-Latex, thick},
  tee/.style={circle, fill=black, inner sep=1.2pt},
  node distance=10mm and 14mm
]

\node[coupler] (c1) {RF Divider\\e.g. Wilkinson};
\node[sdr, right=18mm of c1, yshift=8mm] (sdr1) {SDR\\Receiver 1};
\node[sdr, right=18mm of c1, yshift=-8mm] (sdr2) {SDR\\Receiver 2};

\coordinate (c1in)  at ($(c1.west)+(0,0)$);
\coordinate (c1p1)  at ($(c1.east)+(0,6mm)$);
\coordinate (c1p2)  at ($(c1.east)+(0,-6mm)$);

\draw[line] (c1p1) -- (sdr1.west) node[midway, above] {\scriptsize RX};
\draw[line] (c1p2) -- (sdr2.west) node[midway, below] {\scriptsize RX};

\node[rf, left=10mm of c1] (lna) {LNA};

\node[coupler, left=10mm of lna] (c2) {RF Coupler};

\coordinate (c2p1) at ($(c2.east)+(0,0)$);%
\coordinate (c2p2) at ($(c2.west)+(0,6mm)$);%
\coordinate (c2p3) at ($(c2.west)+(0,-6mm)$);%

\draw[line] (c2p1) -- (lna.west);
\draw[line] (lna.east) -- (c1in);

\node[rf, left=10mm of c2p2] (bpf315) {315 MHz\\Bandpass\\Filter};
\node[rf, left=10mm of c2p3] (bpf433) {433 MHz\\Bandpass\\Filter};

\node[ant, left=10mm of bpf315] (ant315) {Antenna\\315 MHz};
\node[ant, left=10mm of bpf433] (ant433) {Antenna\\433 MHz};

\draw[line] (ant315.east) -- (bpf315.west);
\draw[line] (bpf315.east) -- (c2p2);

\draw[line] (ant433.east) -- (bpf433.west);
\draw[line] (bpf433.east) -- (c2p3);

\node[anchor=south west] at ($(c2p1)+(1mm,1mm)$) {\scriptsize RF Out};
\node[anchor=south east] at ($(c2p2)+(-1mm,1mm)$) {\scriptsize RF1};
\node[anchor=north east] at ($(c2p3)+(-1mm,-1mm)$) {\scriptsize RF2};

\node[anchor=south west] at ($(c1in)+(-4mm,1mm)$) {\scriptsize In};
\node[anchor=south west] at ($(c1p1)+(1mm,1mm)$) {\scriptsize Out 1};
\node[anchor=north west] at ($(c1p2)+(1mm,-1mm)$) {\scriptsize Out 2};

\end{tikzpicture}

  \caption{RF front end block diagram with dual-band coupling and SDR receivers.}
  \label{fig:rf-block}
\end{figure*}
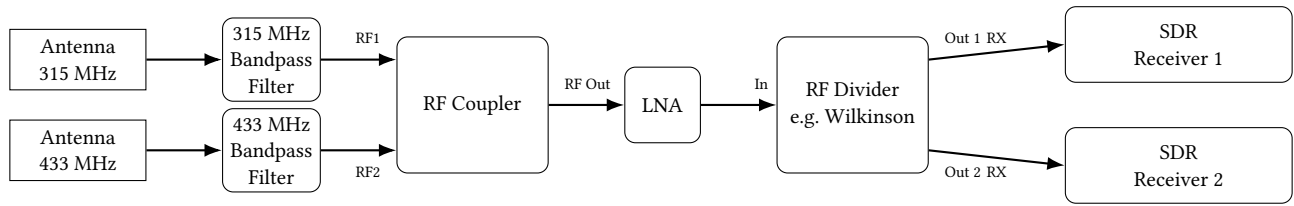

\begin{figure*}[t]
  \centering

\begin{tikzpicture}[
  font=\small,
  >=Latex,
  block/.style={draw, rounded corners, minimum width=10mm, minimum height=10mm, align=center},
  rf/.style={block},
  sdr/.style={block, minimum width=32mm},
  ant/.style={draw, minimum width=20mm, minimum height=8mm, align=center},
  coupler/.style={draw, rounded corners, minimum width=22mm, minimum height=18mm, align=center},
  line/.style={-Latex, thick},
  node distance=10mm and 14mm
]

\node[sdr] (sdr1) {SDR\\Receiver};

\node[rf, left=14mm of sdr1] (lna) {LNA};
\draw[line] (lna.east) -- (sdr1.west) node[midway, above] {\scriptsize RX};

\node[coupler, left=14mm of lna] (c2) {RF Coupler};

\coordinate (c2p1) at ($(c2.east)+(0,0)$);%
\coordinate (c2p2) at ($(c2.west)+(0,6mm)$);%
\coordinate (c2p3) at ($(c2.west)+(0,-6mm)$);%

\draw[line] (c2p1) -- (lna.west);

\node[rf, left=14mm of c2p2] (bpf315) {315 MHz\\Bandpass\\Filter};
\node[rf, left=14mm of c2p3] (bpf433) {433 MHz\\Bandpass\\Filter};

\node[ant, left=14mm of bpf315] (ant315) {Antenna\\315 MHz};
\node[ant, left=14mm of bpf433] (ant433) {Antenna\\433 MHz};

\draw[line] (ant315.east) -- (bpf315.west);
\draw[line] (bpf315.east) -- (c2p2);

\draw[line] (ant433.east) -- (bpf433.west);
\draw[line] (bpf433.east) -- (c2p3);

\node[anchor=south west] at ($(c2p1)+(1mm,1mm)$) {\scriptsize RF Out};
\node[anchor=south east] at ($(c2p2)+(-1mm,1mm)$) {\scriptsize RF1};
\node[anchor=north east] at ($(c2p3)+(-1mm,-1mm)$) {\scriptsize RF2};

\end{tikzpicture}

  \caption{RF front end block diagram with dual-band preselection and a single SDR receiver.}
  \label{fig:rf-block-single-sdr}
\end{figure*}
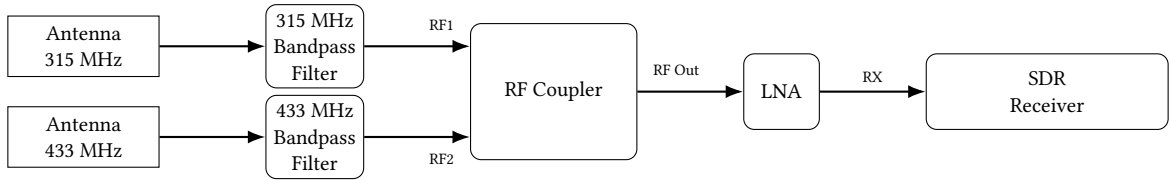

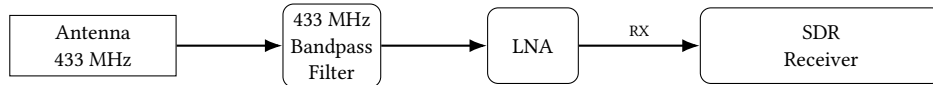
\begin{figure*}[t]
  \centering

\begin{tikzpicture}[
  font=\small,
  >=Latex,
  block/.style={draw, rounded corners, minimum width=12mm, minimum height=10mm, align=center},
  rf/.style={block},
  sdr/.style={block, minimum width=32mm},
  ant/.style={draw, minimum width=22mm, minimum height=8mm, align=center},
  line/.style={-Latex, thick},
  node distance=12mm and 16mm
]

\node[ant] (ant433) {Antenna\\433 MHz};

\node[rf, right=14mm of ant433] (bpf433) {433 MHz\\Bandpass\\Filter};

\node[rf, right=14mm of bpf433] (lna) {LNA};

\node[sdr, right=16mm of lna] (sdr1) {SDR\\Receiver};

\draw[line] (ant433.east) -- (bpf433.west);
\draw[line] (bpf433.east) -- (lna.west);
\draw[line] (lna.east) -- (sdr1.west) node[midway, above] {\scriptsize RX};

\end{tikzpicture}

  \caption{RF front end block diagram with a single 433 MHz antenna, bandpass filter, and LNA placed immediately after the filter feeding a single SDR receiver.}
  \label{fig:rf-block-433-lna-postfilter}
\end{figure*}

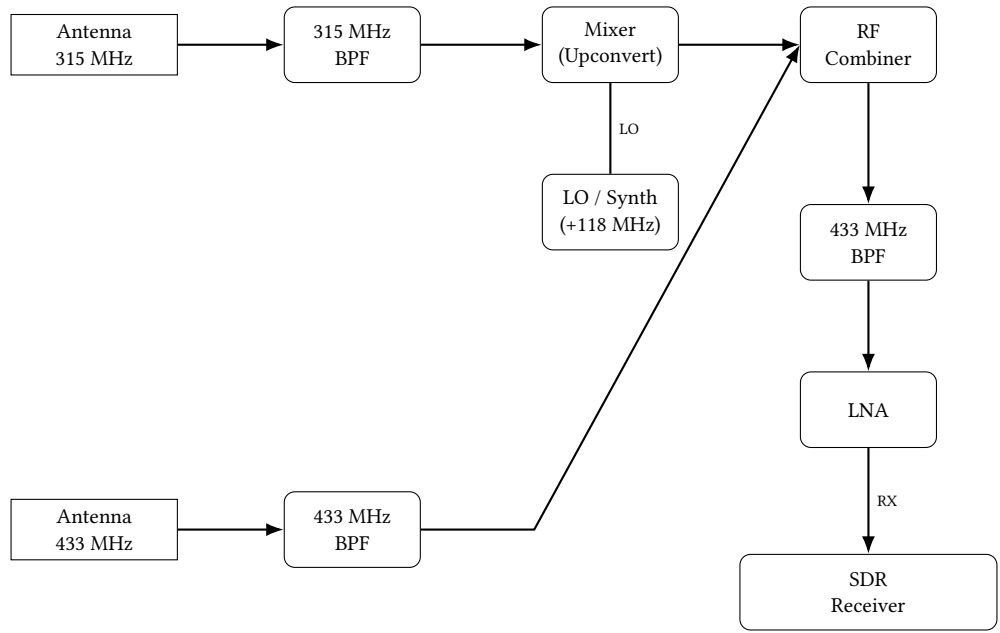
\begin{figure*}[t]
  \centering

\begin{tikzpicture}[
  font=\small,
  >=Latex,
  block/.style={draw, rounded corners, minimum width=18mm, minimum height=10mm, align=center},
  rf/.style={block},
  sdr/.style={block, minimum width=34mm},
  ant/.style={draw, minimum width=22mm, minimum height=8mm, align=center},
  line/.style={-Latex, thick},
  note/.style={font=\scriptsize, align=left},
  node distance=12mm and 14mm
]

\node[ant] (ant315) {Antenna\\315 MHz};
\node[ant, below=56mm of ant315] (ant433) {Antenna\\433 MHz};

\node[rf, right=14mm of ant315] (bpf315) {315 MHz\\BPF};
\node[rf, right=14mm of ant433] (bpf433) {433 MHz\\BPF};

\draw[line] (ant315.east) -- (bpf315.west);
\draw[line] (ant433.east) -- (bpf433.west);

\node[rf, right=16mm of bpf315] (mixer) {Mixer\\(Upconvert)};
\node[rf, below=12mm of mixer] (lo) {LO / Synth\\(+118 MHz)};

\draw[line] (bpf315.east) -- (mixer.west);
\draw[thick] (lo.north) -- (mixer.south) node[midway, right] {\scriptsize LO};

\node[rf, right=16mm of mixer] (comb) {RF\\Combiner};

\draw[line] (mixer.east) -- (comb.west);

\coordinate (tap433) at (bpf433.east);
\coordinate (r1) at ($(tap433)+(15mm,0)$);
\draw[line] (tap433) -- (r1) -- (comb.west);

\node[rf, below=16mm of comb] (bpfout) {433 MHz\\BPF};
\node[rf, below=12mm of bpfout] (lna) {LNA};
\node[sdr, below=14mm of lna] (sdr1) {SDR\\Receiver};

\draw[line] (comb.south) -- (bpfout.north);
\draw[line] (bpfout.south) -- (lna.north);
\draw[line] (lna.south) -- (sdr1.north) node[midway, right] {\scriptsize RX};

\end{tikzpicture}

  \caption{Single-SDR RF front end using upconversion (315 MHz signals are upconverted to the 433 MHz band), eliminating SDR frequency hopping at the cost of band discrimination.}
  \label{fig:rf-block-mixer-vertical}
\end{figure*}

\end{document}